\documentclass{article}
\usepackage{graphicx}
\usepackage{booktabs} % For formal tables
\usepackage{todonotes}
\usepackage{hyperref}

\begin{document}

\title{Protocol-independent Detection of ``Messaging Ordering'' Network Covert Channels}
\author{Steffen Wendzel\\~\\
\small Worms University of Applied Science, Centre of Technology and Transfer (ZTT)\\
\small contact: wendzel@hs-worms.de}
\date{}

\maketitle

\begin{abstract}
\textcolor{red}{This preprint was submitted to CUING'19. The final version of this paper appeared in Proc.\ ARES 2019 and is available here: \url{https://doi.org/10.1145/3339252.3341477}. Enhanced measurements are provided in \cite{Habilthesis}.}\\~\\
Detection methods are available for several known covert channels. However, a type of covert channel that received little attention within the last decade is the ``message ordering'' channel. Such a covert channel changes the order of PDUs (protocol data units, i.e.\ packets) transferred over the network to encode hidden information. The advantage of these channels is that they cannot be blocked easily as they do not modify header content but instead mimic typical network behavior such as TCP segments that arrive in a different order than they were sent.

\textbf{Contribution:} In this paper, we show a protocol-independent approach to detect message ordering channels. Our approach is based on a modified \emph{compressibility score}. We analyze the detectability of message ordering channels and whether several types of message ordering channels differ in their detectability.

\textbf{Results:} Our results show that the detection of message ordering channels depends on their number of utilized PDUs. First, we performed a rough threshold selection by hand, which we later optimized using the C4.5 decision tree classifier.
We were able to detect message ordering covert channels with an accuracy and $F_1$ score of $\geq 99.5$\% and a false-positive rate $<1$\% and $<0.1$\% if they use sequences of 3 or 4 PDUs, respectively. Simpler channels that only manipulate a sequence of two PDUs were detectable with an accuracy and $F_1$ score of $94.5$\% and were linked to a false-positive rate of $5.19$\%. We thus consider our approach suitable for real-world detection scenarios with channels utilizing 3 or 4 PDUs while the detection of channels utilizing 2 PDUs should be improved further.
\end{abstract}


\section{Introduction}

Covert channels are policy-breaking, stealthy communication channels that are reported to be utilized by cyber criminals to enable stealthy communications for malware and data exfiltration~\cite{LucaWojciechIHIntro,mazurczyk2015steganography,panajotov2013covert,CUING}. While traditional steganography methods utilize digital media carriers to embed hidden information (e.g.\ image files or audio files~\cite{Katzenbeisser:2000:IHT:555654}), network covert channels can be realized by hiding data in the meta-data of network traffic, e.g.\ timing of PDUs or modification of PDU content. Covert channels can utilize all OSI layers and were described for all popular communication protocols, e.g.\ DHCP, IPv4, IPv6, ICMP, TCP, UDP, DNS, HTTP, VoIP and others \cite{Wendzel:CSUR:Patterns,zander2007survey,NIHbook,panajotov2013covert,KraetzerDittmann}. The detection of network covert channels is challenging, also for the law-enforcement and for professional administrators, and not all types of covert channels can be efficiently detected.

In recent years, detection approaches improved, rendering trivial covert channels easier to detect. For this reason, covert channels with untypical content, e.g.\ where usually reserved PDU bits are modified, can be detected with simple signatures. In contrast, some covert channels mimic the behavior of normal network traffic~\cite{panajotov2013covert}. A rarely studied type of covert channel are those that can be summarized under the ``message ordering'' pattern~\cite{Wendzel:CSUR:Patterns,NIHbook}. They do not modify the content of network packets but only their order. For instance, the order of TCP segments can be manipulated to encode a hidden message. While the content of the TCP segments would appear normal, the changed order of TCP segments can also be considered normal as such segments do not always arrive in the order they were originally sent. Message ordering channels can be realized by all protocols that can indicate the sequence of packets, e.g.\ TCP, AH and ESP.

In this paper we present an approach to detect message ordering channels in a protocol-independent manner. Our approach is tailored for the law-enforcement to analyze traffic recordings. However, our approach can be potentially applied to corporate networks with a low traffic volume in real-time, too. In particular, the contributions are as follows:
\begin{enumerate}
 \item Perform a modification (\emph{countermeasure variation}) of the so-called \emph{compressibility} score that was originally introduced to detect covert channels that modify the timing between PDUs. Therefore, a suitable coding is presented. The presented detection approach is designed in a way to make it independent of the utilized network protocol.
 \item Evaluating the detectability of message ordering channels that are utilizing a different number of PDUs in their coding.
\end{enumerate}

The remainder of this paper is structured as follows. Sect.~\ref{Sect:FundamentsRelWork} covers fundamentals and discusses related work. Our detection approach and the related countermeasure variation are introduced in Sect.~\ref{Sect:Approach}. We describe our experimental evaluation and discuss our results in Sect.~\ref{Sect:Experiments}. Sect.~\ref{Sect:Concl} concludes.

\section{Fundamentals \& Related Work}\label{Sect:FundamentsRelWork}

\begin{figure}[!h]
\centering
\includegraphics[width=\linewidth]{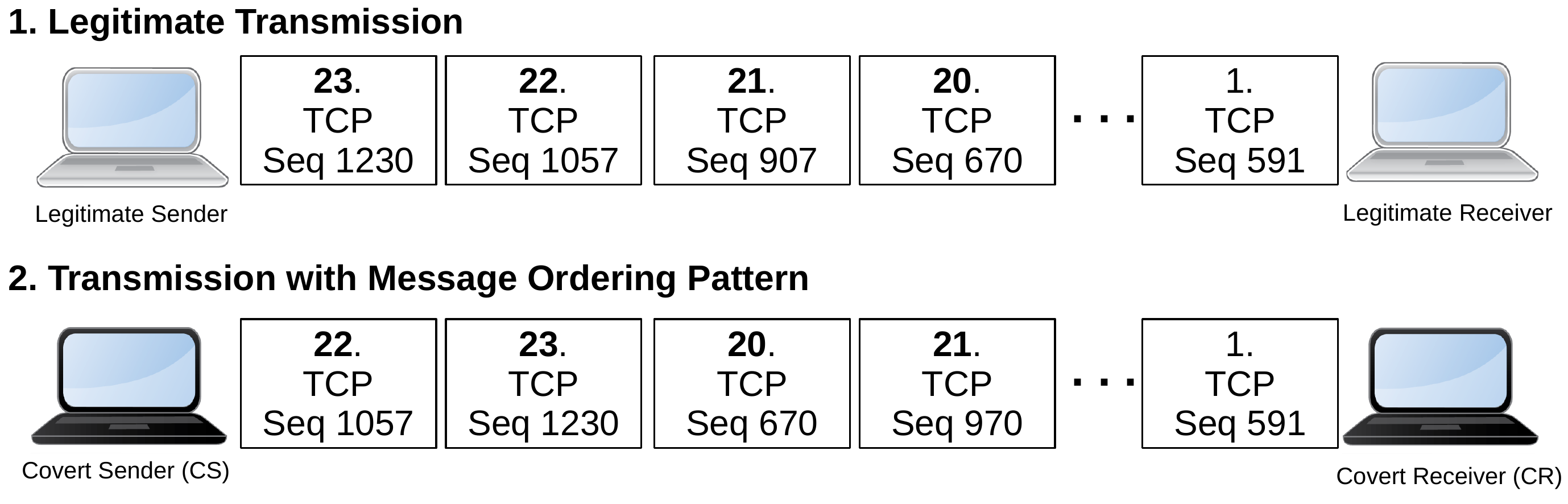}
\caption{The functioning of the message ordering pattern.}
\label{Fig:PDUOrder:Idea}
\end{figure}

%-message ordering channels
The functioning of message ordering channels is visualized in Fig.~\ref{Fig:PDUOrder:Idea}. A covert sender (CS) sends a sequence of messages to the covert receiver (CR). In comparison to legitimate traffic, the order of protocol data units (PDUs), i.e.\ network packets, is modified by a covert sender to encode a hidden message. Sender and receiver agree upon the coding a priori, e.g.\ it can be defined in a malware's executable before deployment.

In general, a covert channel manipulating the order of $n$ packets can be used to transfer $\log_2 n!$ bits \cite{AhsanKundurA,AhsanKundurB}. Thus, the more PDUs are utilized, the more bits can be transferred per PDU. For practicability, longer messages can also be split into $m$ sequences of $n$ PDUs each, allowing $m\cdot \log_2 n!$ bits to be transferred. For instance, if a transfer contains 2,000 packets, and if $n=4$, then this would allow the transfer of $500 \cdot \log_2 4! = 2,292.5$ bits (more than 286 bytes), and it would be 2,762.8 bits (345 bytes) for $n=5$ ($m=400$).
Message ordering channels can utilize the sequence numbers of TCP, AH or ESP headers, several application layer protocol's sequence numbers and the IPv4 Identifier field, among others~\cite{AhsanKundurA,AhsanKundurB,NIHbook}. 

%- Patterns
Hiding patterns are abstract descriptions of covert channel hiding techniques; they were introduced in \cite{Wendzel:CSUR:Patterns}. Each hiding pattern represents a set of covert channel techniques that follow the same general hiding technique. These hiding patterns form a taxonomy, where the message ordering pattern (former called \emph{PDU order pattern}) is categorized as a timing channel (because manipulating the order of PDUs effectively alters their timings). In \cite{NIHbook}, the message ordering pattern was additionally categorized as \emph{protocol-aware}, i.e.\ as a method \emph{that require[s] the understanding of the carrier protocol}\footnote{In contrast, protocol-agnostic timing channels can modify the carrier blindly, e.g.\ by manipulating the inter-arrival times between packets.}, and was finally renamed to \emph{(manipulated) message ordering} pattern.

%- CC Detection
Covert channel detection is also well covered in the existing literature. In general, countermeasures either aim on detecting the presence of the channel itself or the involvement of a participant in the covert communication. Other countermeasures aim on limiting the channel capacity or on preventing the use/the existence of channels. Moreover channels can be audited \cite{NIHbook}. Several detection methods exist and summaries can be found in \cite{zander2007survey,NIHbook}. A recent trend is the detection of distributed covert channels \cite{Cabaj:2018:TDN:3230833.3233264}. Message ordering channels could also be realized in a distributed manner, e.g.\ split over multiple flows simultaneously. We do not explicitly focus on such a distributed scenario in this paper, but the presented approach could potentially detect distributed message ordering channels nevertheless on a per-flow basis. 
Moreover, digital media steganography has similar channels, e.g.\ methods that modify the order of HTML tags in websites, which can be detected with statistical methods~\cite{HTMLTagStegoDetect}. However, no specific method was presented so far to detect network-based message ordering channels.

%- Compressibility Score
A detection approach for covert timing channels it the so-called \emph{compressibility score} that was introduced by Cabuk et al.~\cite{CabukEtAlPaper}. The compressibility score is applied to detect channels that modulate inter-arrival times (IAT) between succeeding network packets. These channels belong to the ``Inter-packet Times'' pattern~\cite{NIHbook}. Therefore, IAT values of a flow are recorded, and then rounded. The rounded values are encoded as short strings that are concatenated to a long string $S$. Afterwards, $S$ is compressed with a compressor $\Im$ (e.g.\ \emph{gzip}), i.e.\ $C=\Im(S)$. Finally, the value $\kappa = |C| / |S|$ is calculated. $\kappa$ is used as an indicator for the presence of a covert timing channel. In general, a covert channel's artificial IAT values of the ``inter-packet times'' pattern would be similar within a given flow, rendering the results better compressible than legitimate traffic with more varying IAT values~\cite{CabukEtAlPaper}. In comparison to the inter-packet times pattern, the message ordering pattern does not encode hidden information in the IAT values but in the pure order of PDUs, i.e.\ it does not matter how large the IAT values between PDUs are.

%- Countermeasure Variation
The concept of countermeasure variation was recently introduced in \cite{wendzel2018one}. The idea is to take existing countermeasures, e.g.\ detection approaches as the one of Cabuk et al., and modify them slightly so that they can work for other patterns as originally intended. Countermeasure variation aids the problem of not having countermeasures for all patterns available. Previous work has shown that the compressibility score as well as the so-called $\epsilon$-similarity\footnote{The $\epsilon$-similarity is described in \cite{cabuk2004ip,CabukEtAlPaper}.} can be adjusted to work with the ``size modulation''~\cite{wendzel2018one} and ``artificial re-transmissions'' patterns~\cite{NordSec18}. In this paper, we perform a countermeasure variation of the compressibility score so that it can work with the message ordering pattern.

\section{Detection Approach}\label{Sect:Approach}

The development of countermeasures for criminal uses of covert channels and steganography has recently attracted increasing attention, cf.~\cite{LucaWojciechIHIntro,CUING}. Therefore, we consider a scenario where traffic recordings of criminal cases must be analyzed some time after they were recorded. This means that we do not focus on a real-time detection scenario, i.e.\ computing performance of our approach is not a strong requirement. However, it can potentially be applied in low-volume networks, too, for real-time detection.

Our detection approach is designed to work for intermediate nodes (IM) located between covert sender (CS) and covert receiver (CR) that conduct traffic recordings. It can be a gateway or router. However, as an administrative system process, our detection approach could also be integrated into CS or CR as long as the network covert channel process and its user are separated from the detection process. This could be realized, e.g.\ in the form of a malware. However, in this case, the detection process needs direct access to the network interface, e.g.\ as a kernel-level routine. As it is beneficial to recognize \emph{all} packets of a flow, it would even be optimal to locate the detection process on CS or CR since there might be multiple IMs for multiple routing paths, so one IM might not see \emph{all} traffic exchanged between CS and CR.

We perform a countermeasure variation as follows. First, we modify the compressibility score of Cabuk et al.\ in a way that we provide it with a different input as for the algorithm. Second, we modify the generation process of the string $S$. Third, we determine optimal thresholds to differentiate between legitimate and covert channel traffic. The other steps of the compressibility score (compression and calculation of $\kappa$) remain as in the original approach. The modifications are explained in subsection~\ref{Subsect:AlgoInputAndStringEncoding}.

\subsection{Algorithm Input and String Coding}\label{Subsect:AlgoInputAndStringEncoding}
In the original approach by Cabuk et al., only one algorithm is shown to encode IAT values into strings. Instead, we encode sequence numbers and moreover experimented with four different approaches to encode these sequence numbers. Another difference to the original approach by Cabuk et al.\ is that IAT values cannot overrun but sequence numbers can indeed overrun. For this reason, we implemented a filter. We also defined new detection thresholds for $\kappa$.

\begin{figure}[!h]
\centering
\includegraphics[width=\linewidth]{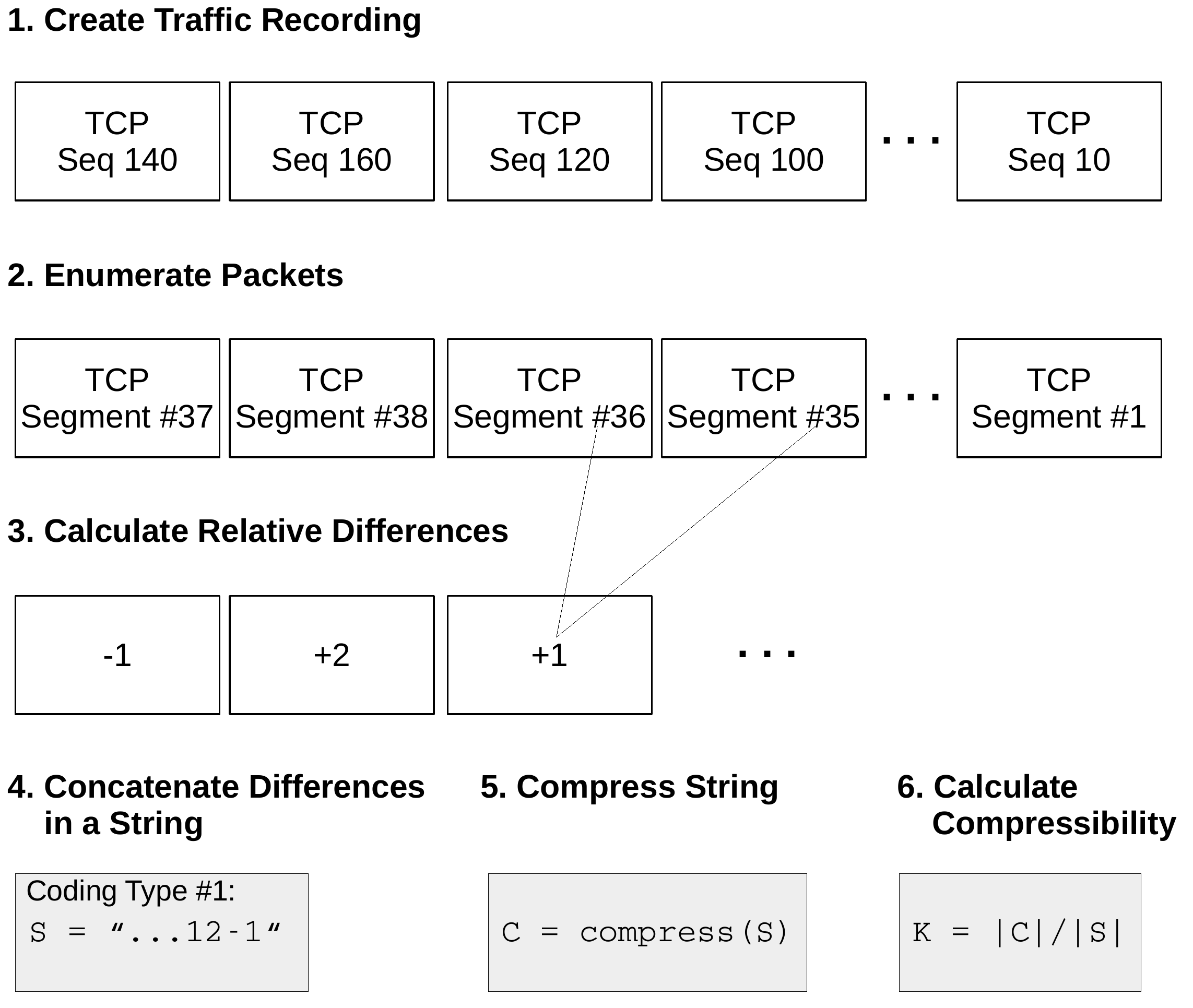}
\caption{Our proposed five-step approach to detect message ordering channels (steps 3 and 4 are exemplified for one type of coding).}
\label{Fig:process}
\end{figure}

The details of our approach are described in the following. Fig.~\ref{Fig:process} visualizes our approach.

In step 1, we record all flows between two or more hosts to be observed (Fig.~\ref{Fig:process}-1). For each flow, we consider a window of 200 PDU sequences, i.e.\ 201 packets. For instance, such a flow could contain 201 TCP segments (or alternatively a flow using any other protocol with a header field that allows to determine the PDU order).\footnote{Please note that the message ordering pattern requires protocols to have a numbering element. Thus, protocols without such a field are not of relevance for our research.} For all these sequences, we assign each PDU a number between 1 and 200 based on its appearance in the order of packets in the window.
For instance, if four following sequence numbers were 100, 120, 160, 140, for the packets 35 to 38, then we would record the order 35, 36, 38, 37 (see Fig.~\ref{Fig:process}-2).

In the following, \texttt{diff} represents the relative difference between the current and the last PDU number. For the mentioned example, the values for \texttt{diff} would be +1, +2, and -1. \texttt{|x|} indicates the absolute value of the variable \texttt{x}. Moreover, \texttt{a || b} is the string concatenation of the strings \texttt{a} and \texttt{b}. We use \texttt{a || (y ? 't' : 'f')} to test whether the condition \texttt{y} is true. If it is true, the string \texttt{'t'} will be concatenated to the string \texttt{a}, otherwise the string \texttt{'f'} will be concatenated to \texttt{a}. The function \texttt{str(x)} returns the value of \texttt{x} in the form of a string.

We tested the following four different codings. Please note that only one string element is shown each, i.e., 200 of these values would be concatenated to create $S$.
%1. diff
%2. abs(diff)
%3. diff || (abs(diff) % 2 ? A : B)
%4. skip(olddiff>diff*5); abs(diff) || (abs(diff) % 2 ? A : B)     <-- best
\begin{enumerate}
 \item \texttt{str(diff)}: concatenate the \texttt{diff} values of succeeding packets (exemplified in Fig.~\ref{Fig:process}).
 \item \texttt{str(|diff|)}: concatenate the absolute \texttt{diff} values of succeeding packets.
 \item \texttt{str(diff) || ( |diff| \% 2 ? 'A' : 'B' )}: for all succeeding packets, concatenate \texttt{diff} with an `A' if the absolute value of \texttt{diff} $\bmod$ 2 equals 1, otherwise with `B'.
 \item \texttt{str(|diff|) || ( |diff| \% 2 ? 'A' : 'B' )}: for all succeeding packets, concatenate the absolute value of \texttt{diff} with an 'A' if the absolute value of \texttt{diff} $\bmod$ 2 equals 1, otherwise with 'B'.
\end{enumerate}

Fig.~\ref{Fig:process}-3 and 4 exemplify coding type 1, i.e.\ a simple concatenation of the \texttt{diff} values. However, in result, coding type 4 performed best using our training data-set. Using coding 4, a perfect transmission without retransmissions and with no out-of-order packets would look like ``\texttt{1A1A1A1A1A1A$\ldots$1A1A1A}'' with $|S|=400$, resulting in $|C|=27$ and thus $\kappa=14.81481$ (two characters per difference). Indeed, we observed that several legitimate flows (approx.~28\%) had exactly this $\kappa$ value and the related string generate from 200 sequence numbers each.

%This is complicated, but should be correct:
%XXX Overall: improve terminology use ("PDU order number" etc.).
Sometimes, sequence numbers overrun. For instance, if a sequence number field has eight bits, the sequence number following 255 would be 0. These overruns are rare, but we filtered them by checking whether the difference in the current PDU number is larger than 5 times the previous PDU number difference (\texttt{olddiff}) using the expression \texttt{olddiff > diff*5}. 
%Hää?
%This expression usually filters two PDU number differences as sequence numbers directly influence the PDU number differences.
%End:Hää?
This filter slightly improved detection results additionally. Moreover can this filter handle typical sequence number fields. % as long as the difference of 5 can be reached.
%\footnote{However, if the sequence number field would have only 1-2 bits, our filter would not work.}
We tested the algorithm with TCP (32 bit sequence numbers) and the CCEAP covert channel\footnote{\url{https://github.com/cdpxe/CCEAP/}} tool (8 bit sequence numbers).

Finally, we compressed the concatenated string $S$ using $C=\Im(S)$ and then calculated $\kappa=|C|/|S|$ as done in the original approach by Cabuk et al.\ (Fig.~\ref{Fig:process}-5 to \ref{Fig:process}-6).

\subsection{Parameterization}

\begin{figure}[h]
  \centering
  \includegraphics[width=\linewidth]{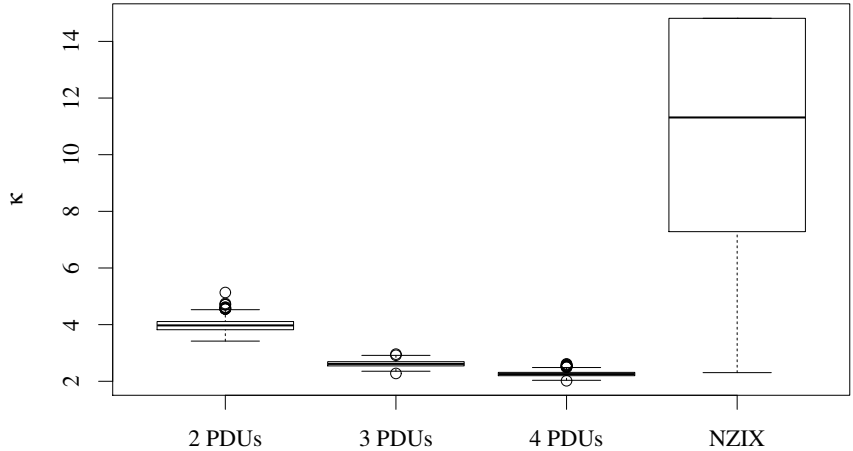}
  \caption{\label{Fig:TrainAnalysis}Analysis of NZIX training data in comparison to covert channels using sequences of 2 and 4 PDUs.}
\end{figure}

For each a flow, in order to decide whether a covert channel or a legitimate data flow is present, a threshold for the value $\kappa$ must be defined. Therefore, we extracted 454 flows with at least 201 sequence numbers from the NZIX data set\footnote{\url{ftp://wits.cs.waikato.ac.nz/pma/long/nzix/2/}} and calculated their compressibility scores. Please note that the NZIX data are coming from a real Internet environment and contain PDUs which are reordered due to legitimate reasons. We compared the NZIX scores with the compressibility scores of three different covert channel types, modifying the order of 2 to 4 PDUs, both randomly to represent the transmission of encrypted content. As shown in Fig.~\ref{Fig:TrainAnalysis}, all types of covert channels resulted in comparably $\kappa$ values: channels with 2 PDUs had a mean of 3.983 while those with 3 PDUs had a mean of 2.621 and those with 4 PDUs had a mean of 2.259. The NZIX training data resulted in a mean $\kappa$ value of 10.674. However, as visible in Fig.~\ref{Fig:TrainAnalysis}, there is a clear overlapping of compressibility results between especially NZIX training data and covert channels using 2 PDUs. For this reason, we decided to parametrize our detection approach with thresholds in the range of $\kappa = \langle 2 ; 14\rangle$\footnote{The maximum achievable compression using \texttt{gzip} was slightly above 14.8, thus rendering higher values insignificant. For compression, we write the string $S$ into a file and run \texttt{gzip -9 --no-name <filename>} to calculate $C$.}, while the range $\langle 2.5 ; 5.5 \rangle$ was studied in detail since only few covert channel compressibility scores were above $5.5$.

\section{Experiments}\label{Sect:Experiments}

We used the aforementioned tool CCEAP to create message ordering covert channels. Therefore, we extended the tool so that randomized sequence numbers that represent encrypted content could be created.
 We generated between 100 and 4000 pairs of 2, 3 and 4 PDU each, %, using time differences of 10ms, 20ms, 50ms, 100ms, 200ms and 500ms,
 and recorded the generated traffic. Each flow configuration was repeated 20 times and with different inter-arrival times between 10ms and 500ms. Overall, we created 2,160 covert channel flows (720 for every 2, 3 and 4 PDU channel). We extracted the sequence numbers as described in Sect.~\ref{Sect:Approach} and calculated the compressibility scores for all these flows.

Afterwards, we extracted flows with at least 200 sequence numbers from NZIX traffic recordings, resulting in approx.~8,500,000 packets (approx.~ 1,100 flows).

Next, we compared the detectability of every CCEAP configuration with the NZIX recordings. Each time, the same number of 720 legitimate and 720 covert channel flows were used to calculate the detectability. Thus, for all thresholds (2, 2.5, 2.75, 3, 3.25, 3.5, 3.75, 3.9, 4, 4.025, 4.05, 4.075, 4.1, 4.15, 4.2, 4.25, 4.3, 4.4, 4.5, 4.6, 4.7, 4.8, 4.9, 5, 5.5, 6, 7, 8, 9, 10, 11, and 12) we measure the detectability for message ordering channels utilizing 2, 3 and 4 PDUs.

The measurement of the detectability was based on the metrics precision, recall, accuracy and $F_1$-score and all these measures are visualized in the following figures. \emph{Precision} is amount of flows correctly classified as covert channel flows (true positives, TP) in comparison to all flows classified as covert channels (TP and false positives, FP) while \emph{recall} is the number of correctly classified covert channel flows in comparison to the correctly classified covert channel flows plus those flows that were classified as legitimate traffic but were actually covert channel flows, too (false negatives, FN).
\begin{eqnarray*}
\mathrm{precision} &=& \frac{\mathrm{TP}}{\mathrm{TP+FP}}, \\
\mathrm{recall} &=& \frac{\mathrm{TP}}{\mathrm{TP+FN}}.
\end{eqnarray*}
\emph{Accuracy} is the number of true classifications (positive and negative, i.e.\ TP and TN) in comparison to the whole population of flows. Finally, the \emph{F$_1$-score} is the harmonic mean of precision and recall.
\begin{eqnarray*}
\mathrm{accuracy} &=& \frac{\mathrm{TP+TN}}{\mathrm{TP+TN+FP+FN}}, \\
\mathrm{F}_1\mathrm{-score} &=& \frac{2 \cdot \mathrm{precision} \cdot \mathrm{recall}}{\mathrm{precision+recall}}.
\end{eqnarray*}

\begin{table}
 \begin{tabular}{c|cc}
    \toprule
    \textbf{Seq.~Items} & \textbf{Accuracy} & \textbf{$F_1$ Score} \\
%
% OUTPUT OF results_to_tex_table.sh STARTS HERE
%
\midrule
$\kappa=2.5$:\\
2     &	49.722\%	& 0\% \\
3     &	55.833\%	& 21.673\% \\
4     &	99.027\%	& 99.022\% \\
\midrule
$\kappa=2.75$:\\
2     &	49.513\%	& 0\% \\
3     &	92.847\%	& 92.375\% \\
\textbf{4}    &	\textbf{99.513\%} & \textbf{99.516\%} \\
\midrule
$\kappa=3$:\\
2     &	49.236\%	& 0\% \\
\textbf{3}  & \textbf{99.236\%} & \textbf{99.241\%} \\
4     &	99.236\%	& 99.241\% \\
\midrule
$\kappa=3.25$:\\
2     &	48.958\%	& 0\% \\
3 \& 4&	98.958\%	& 98.968\% \\
\midrule
\midrule
$\kappa=4.25$:\\
2     &	90.555\%	& 90.38\% \\
3 \& 4&	96.18\%	& 96.32\% \\
\midrule
$\kappa=4.3$:\\
2     &	91.805\%	& 91.815\% \\
3 \& 4&	95.833\%	& 95.999\% \\
\midrule
$\kappa=4.4$:\\
2     &	93.333\%	& 93.494\% \\
3 \& 4&	95.416\%	& 95.617\% \\
\midrule
$\kappa=4.5$:\\
2     &	94.166\%	& 94.384\% \\
3 \& 4&	95.138\%	& 95.364\% \\
\midrule
$\kappa=4.6$:\\
\textbf{2}    &	\textbf{94.513\%}	& \textbf{94.756\%} \\
3 \& 4&	94.93\%	& 95.174\% \\
%
% OUTPUT OF results_to_tex_table.sh ENDS HERE
%
    \midrule
    \midrule
    C4.5 Classifier: \\
    %   accuracy  f1-score
    2 & 95.9\% (+1.39\%)  & 95.9\% (+1.14\%) \\ 
    3 & 99.54\% (+0.30\%) & 99.55\% (+0.31\%) \\ % micro tree <=2.88659 and > that value
    4 & 99.84\% (+0.33\%) & 99.8\% (+0.28\%) \\ % micro tree <=2.59047 and > that value
    \bottomrule
 \end{tabular}
\caption{\label{Tab:DetectionResultsPerCCType}Detection results, depending on the number of sequence items for selected thresholds (improvement of C4.5 classifier over best manually selected threshold).}
\end{table}

\subsection{Experimental Results}

\begin{figure}[ht]
\centering
\includegraphics[width=\linewidth]{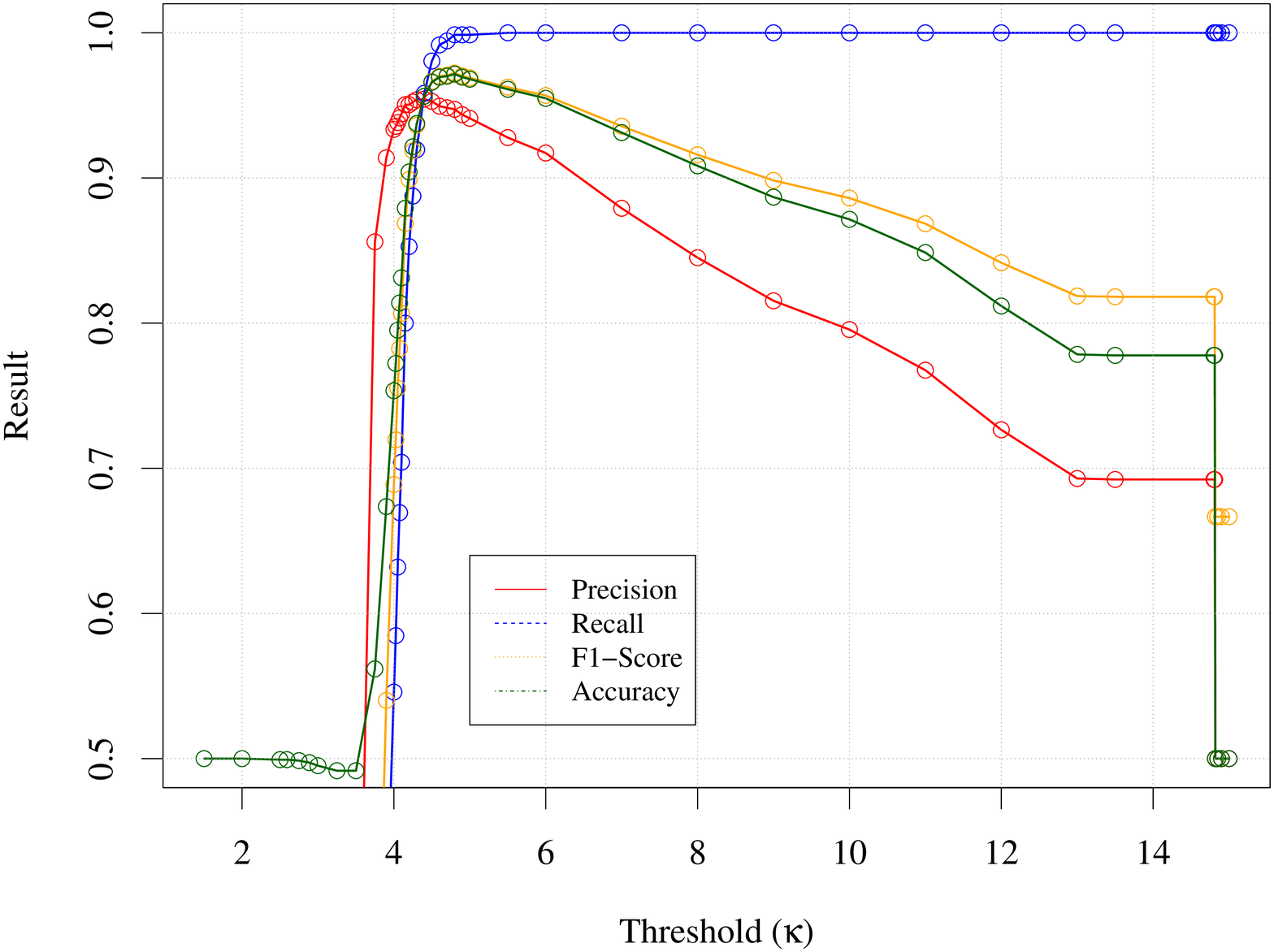}
\caption{Detection results for 2-PDU message ordering channels.}
\label{Fig:DetectionResults:2}
\end{figure}

First, we tested all parameters for covert channels using two symbols. The results are shown in Fig.~\ref{Fig:DetectionResults:2}. Here, the threshold of $\kappa=4.6$ performed best (also see Tab.~\ref{Tab:DetectionResultsPerCCType} for detailed results). However, the achieved accuracy of 94.513\% and $F_1$ score of 94.756\% are not satisfying for large volumes of traffic. The thresholds $\kappa<4.4$ and $\kappa \geq 4.9$ were not leading to acceptable results (accuracy and $F_1$ score below 94\%).

\begin{figure}[ht]
\centering
\includegraphics[width=\linewidth]{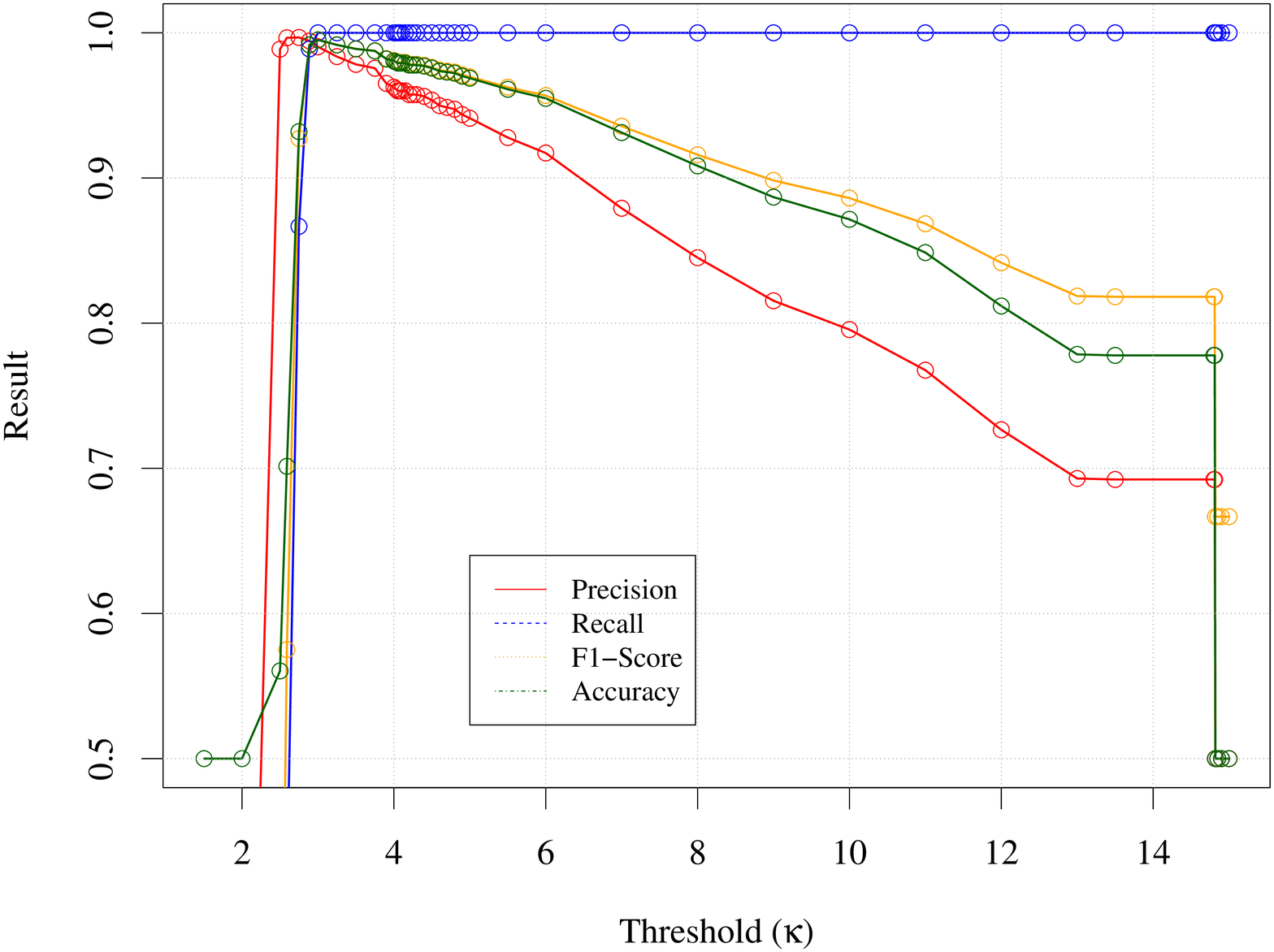}
\caption{Detection results for 3-PDU message ordering channels.}
\label{Fig:DetectionResults:3}
\end{figure}

As shown in Fig.~\ref{Fig:DetectionResults:3}, the optimal threshold for channels that were utilizing 3 PDUs appears to be better for lower values of $\kappa$ than in case of channels utilizing 2 PDUs. For the 3 PDU channels, acceptable values (both, accuracy and $F_1$ score $\geq$94\%) were achieved for $\kappa=3$ to $\kappa=4.8$ with the best results being achieved for $\kappa=3$ ($\geq$99.23\% accuracy and $F_1$ score).

\begin{figure}[ht]
\centering
\includegraphics[width=\linewidth]{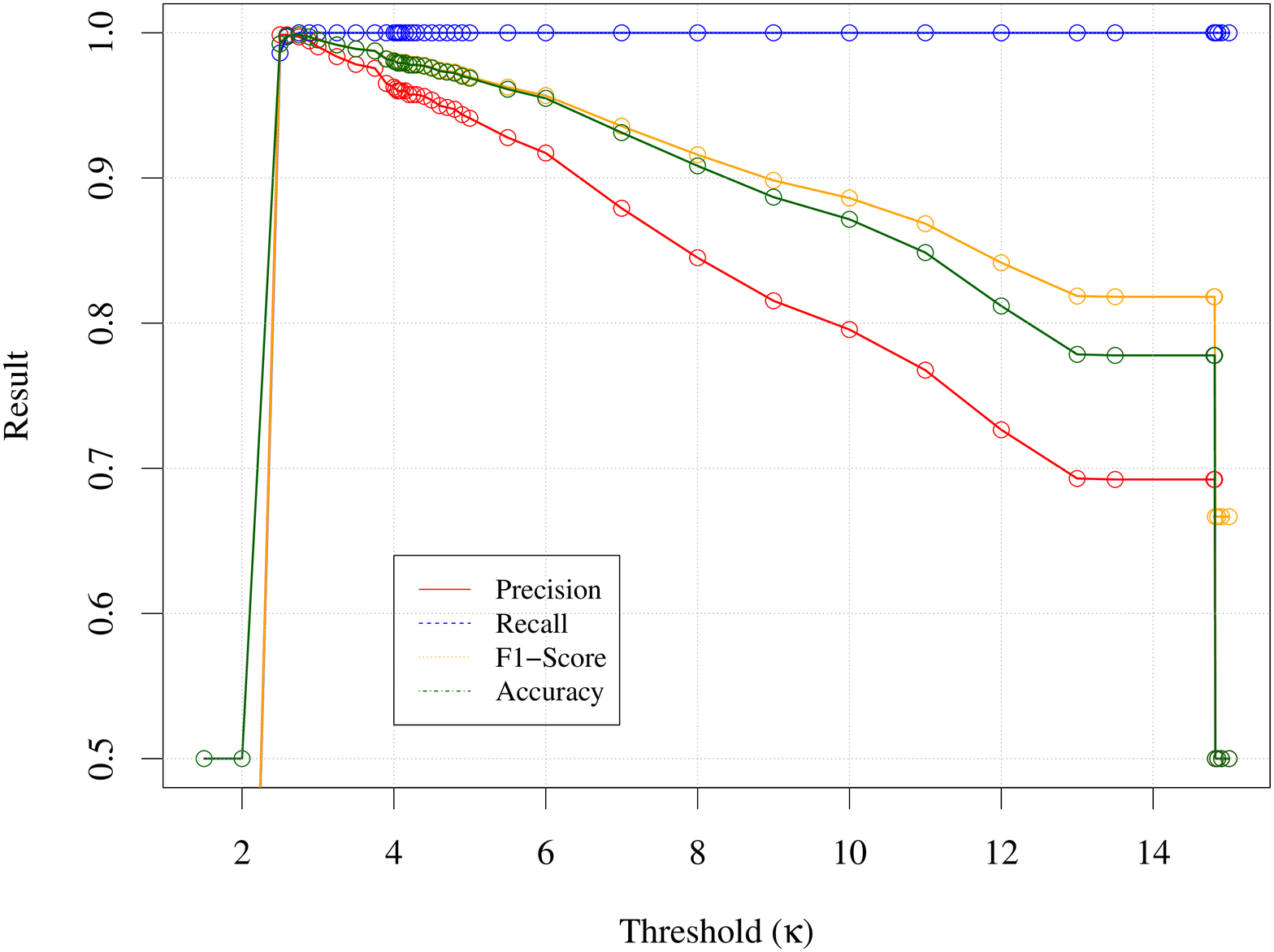}
\caption{Detection results for 4-PDU message ordering channels.}
\label{Fig:DetectionResults:4}
\end{figure}

For a more sophisticated channel that would utilize 4 symbols, the optimal threshold of $\kappa=2.75$ was again lower (Fig.~\ref{Fig:DetectionResults:4} and Tab.~\ref{Tab:DetectionResultsPerCCType}). Results for $\kappa \geq 3$ were equal to channels utilizing 3 PDUs as the maximum compressibility scores of both channels were each below the same maximum value. Acceptable results of $\geq$94\% accuracy and $F_1$ score were achieved for $2.5 \leq \kappa \leq 4.8$.

\begin{figure}[h]
  \centering
  \includegraphics[width=\linewidth]{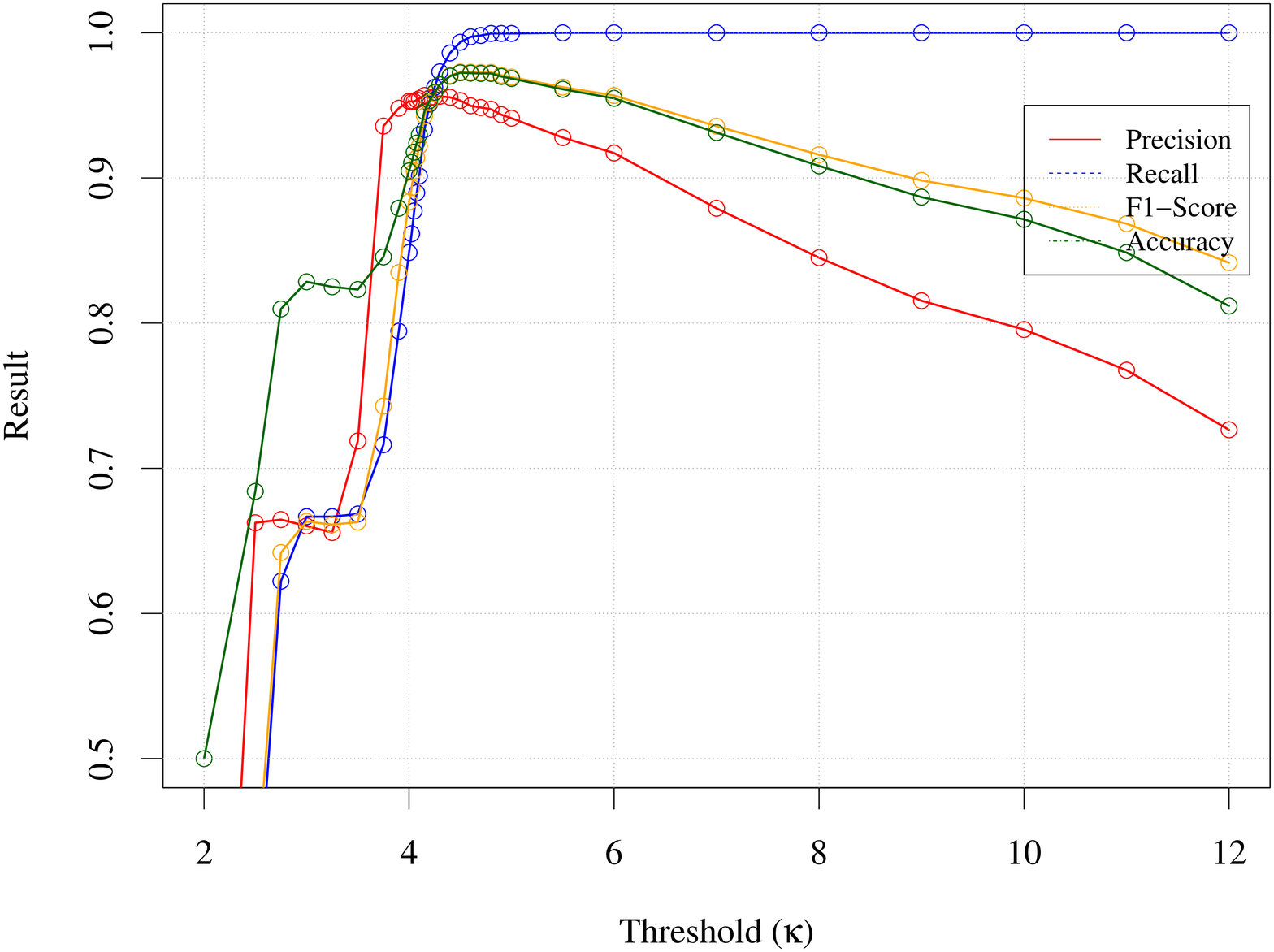}
  \caption{\label{Fig:DetectionResults:Heuristic:All}Detection results for a mixture of the three previously used covert channel types.}
\end{figure}

Finally, we investigated how well a mixture of covert channels utilizing 2, 3 or 4 PDUs can be detected. Therefore, we combined the legitimate NZIX flows with covert channel flows utilizing 2, 3 and 4 PDUs. We used 50\% NZIX flows and 50\% covert channel flows, of which 1/3 were using the same number of PDUs each.
Fig.~\ref{Fig:DetectionResults:Heuristic:All} shows that the optimal threshold appears to be approximately around $\kappa=4.5$. However, differences between $\kappa=4.4$ to $\kappa=4.7$ were only marginal (Tab.~\ref{Tab:AllCCMixedResults}), indicating that the threshold is value is not highly sensitive.

\subsection{Consideration of Realistic Conditions}

Under realistic conditions, we can expect the percentage of covert channel traffic using the message ordering pattern to be very low. If we assume that amount of covert channel traffic is close 0\%, then an important factor to be considered is the \emph{false-positive rate} (FPR), i.e.\ the amount of legitimate traffic that was considered to be covert channel in comparison to all legitimate traffic:
\begin{eqnarray*}
\mathrm{FPR} &=& \frac{\mathrm{FP}}{\mathrm{FP+TN}}.
\end{eqnarray*}

If the FPR is high, the number of false-positives could easily lead to extensive analysis work in an operational network.

\begin{figure}[ht]
\centering
\includegraphics[width=\linewidth]{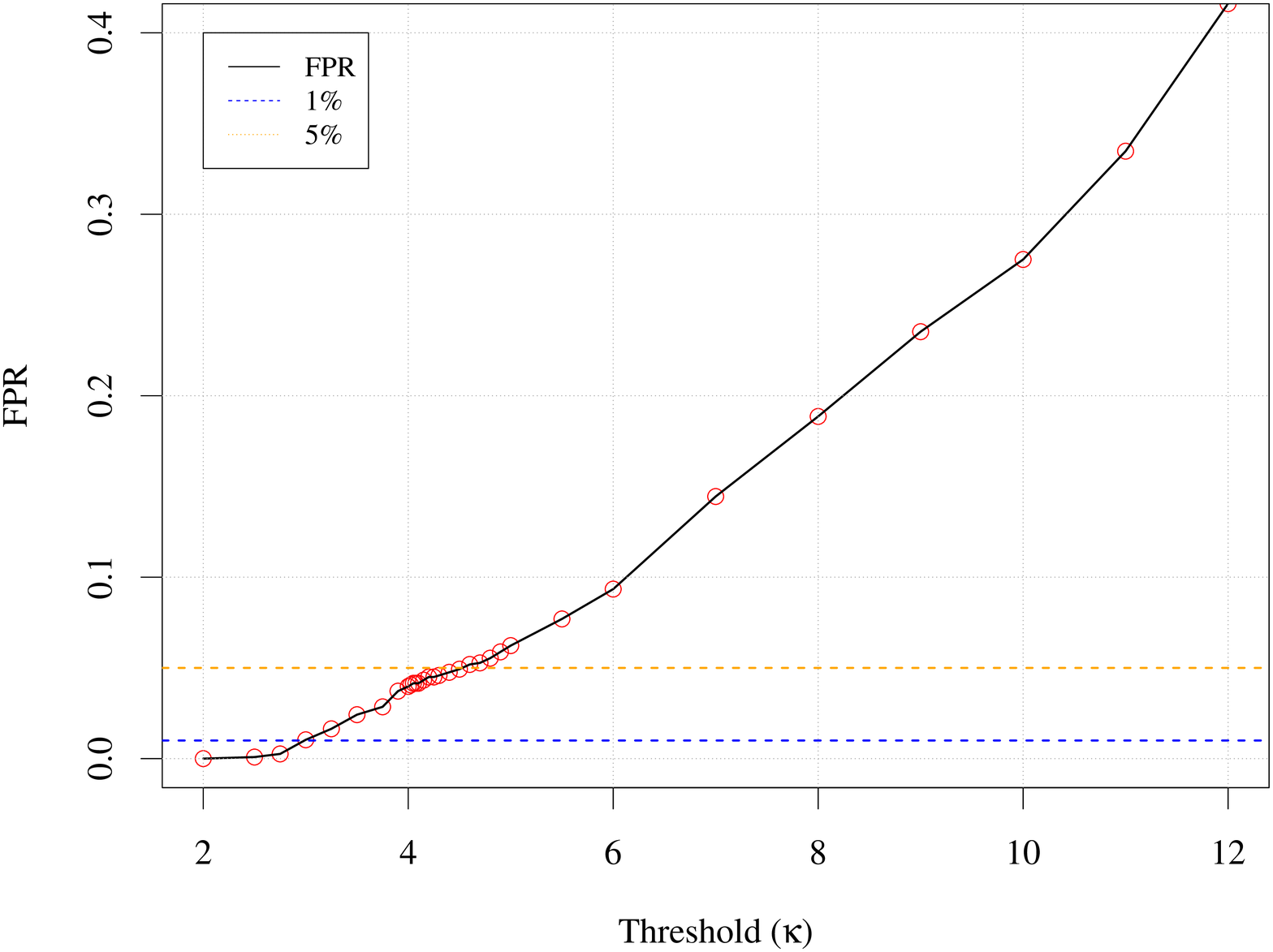}
\caption{False-positive rate for our threshold-based detection.}
\label{Fig:FPR}
\end{figure}

We thus analyzed the FPR of legitimate-only traffic for different thresholds. Therefore, we analyzed 1,156 NZIX flows with $\geq 200$ sequence numbers extracted from one split PCAP file with approximately 8,750,000 packets. The results are shown in Fig.~\ref{Fig:FPR}. The FPR increases to more than 5\% for thresholds $\kappa > 4.5$, rendering our approach impractical for these thresholds. However, for the thresholds 2 to 3.9 the FPR is between 0\% and 3.719\%.

If the optimal thresholds of $\kappa=2.75$ to $\kappa=3.0$ for channels using 3 or 4 PDUs are selected, the FPR is kept at 0.259\% and 1.038\%.

However, to detect channels that utilize only 2 PDUs, we optimally apply a threshold of $\kappa=4.6$. This threshold leads to a FPR of 5.19\% and thus would be impractical to handle in real-world scenarios as discussed in \cite{Steinebach:CUING18Paper}. The overall optimal threshold for a \emph{mixture} of covert channel types ($k=4.5$) results in a FPR of 4.93\%.

We thus conclude that our detection approach can be only considered practical under the following conditions:

\begin{itemize}
\item The detection of channels utilizing 2 or more PDUs is practical for very low traffic volumes such as in small enterprises or office networks due to the rather high FPR of 4.93\%.
\item The detection of channels utilizing 3 or more PDUs is practical in real-world scenarios in combination with a low-threshold due to a FPR of 0.259\% and 1.038\%. We expect that channels utilizing $\geq 5$ PDUs are also providing low compressibility scores and should thus be detectable even easier with our approach.
\end{itemize}

In other words, channel types with the largest capacity can be detected the best while channels with the lowest capacity (utilizing only 2 PDUs) can be considered difficult to detect with this approach.

\subsection{Further Optimization Using C4.5 Classifier}

We trained the J48 classifier, which is an implementation of the C4.5 decision tree algorithm, using the tool \texttt{weka} with only one feature, i.e.\ the compressibility score of flows. We used only this single feature as no other traffic feature would directly be influenced by the message ordering pattern, potentially despite the inter-arrival time of packets. However, even with only one feature, C4.5 can be used for two purposes:

\begin{enumerate}
\item To compare our selected threshold's performance with one or multiple thresholds automatically selected by C4.5. In this sense, we applied C4.5 as a means of quality control for our threshold-based detection approach. Our threshold-based approach provided good detection results but if our threshold was selected close to the optimum value, then its result should be only slightly lower than those of the C4.5 classifier that uses the same feature.\footnote{This is rooted in the fact that the classifier cannot only automatically determine a suitable threshold but can also divide the selection of legitimate and covert channel flows by applying several thresholds in a row.}
\item To let C4.5 find suitable thresholds that we can later use for our threshold-based selection instead of our previously selected and discussed thresholds. Applying these new thresholds will further allow us to calculate the FPR.
\end{enumerate}

First, we studied all three covert channel types separately. With 660 NZIX and 660 covert channel flows (each, i.e.\ per number of utilized PDUs) and a 10-fold cross-validation, we achieved a slightly higher detectability than using the previous thresholds. C4.5 resulted in slightly different thresholds for $\kappa$ and most decision trees had only one decision element. 
For a channel with 3 symbols, the C4.5-selected $\kappa$-threshold was 2.88659 (previously 3.00), which resulted in a FPR of 0.605\% (instead of the previous 1.038\%). 
For a channel with four PDUs, the threshold for $\kappa$ was 2.59047 (previously 2.75) and the linked FPR was even below 0.1\% (0.086\% instead of the previous 0.259\%). This means that both thresholds were close to the original thresholds but improved the FPR values further.

The results increased even more for covert channels using two symbols (more than one percent improvement in accuracy and $F_1$ score). However, in comparison to the other two channel types, 2-PDU channels resulted in a larger decision tree and our results came from an non-pruned tree (over-fitted) and would decrease to the previous level after pruning.\footnote{The major threshold selected by C4.5 for the 2-PDU channel was 4.6 (i.e., exactly the one that we already selected).}

\begin{figure}[ht]
\centering
\includegraphics[width=\linewidth]{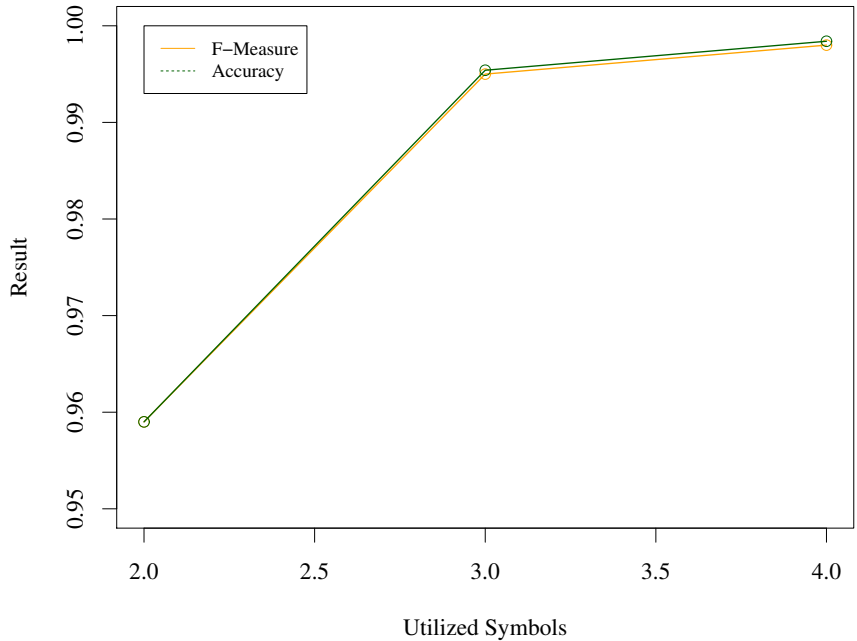}
\caption{$F_1$ Score and Accuracy for D4.5, depending on the number of utilized symbols.}
\label{Fig:C45:DetectionResults:All}
\end{figure}

Fig.~\ref{Fig:C45:DetectionResults:All} visualizes the C4.5-based detectability for all 3 covert channel types depending on the number of utilized PDUs. For exact comparison, the results are also shown at the end of Tab.~\ref{Tab:DetectionResultsPerCCType}. Overall, the new thresholds provided an improvement of approx.\ 0.3\% for accuracy and $F_1$ score of channels utilizing 3 or 4 PDUs.

Finally, we compared the overall detectability of the threshold-based approach with the C4.5 classifier for all three covert channel types. As shown in Tab.~\ref{Tab:AllCCMixedResults}, the C4.5 classifier outperformed our simpler threshold-based detection approach that used $\kappa=4.5$ (improvement of 1.91\% in accuracy and 1.66\% in $F_1$ score). Again, we applied a non-pruned tree here to determine a maximum value. We can conclude that there is still slight room for improvement when it comes to the detection of a mixture of the three covert channels. However, even the detection results of C4.5 with an over-fitted tree were comparable to our simpler threshold-based detection approach.

% Accuracy + F_1 Score:
% XXX diese Werte immer neu berechnen, wenn ich was ändere und dann in Tabelle
% 	unten drunter übernehmen!
%wendzel@sw-virtMate:~/svn/papers/CUING19_SeqModPaper$ cat detection_results.db | awk -F\; 'BEGIN{acc_sum=0;f1_sum=0;counter=0;}{if ($1 == "4.4"){acc_sum+=$9;f1_sum+=$10;counter+=1;}}END{print "counter=" counter ", accuracy="100*acc_sum/counter, ", f1-score="100*f1_sum/counter;}'
%counter=3, accuracy=94.7217 , f1-score=94.9093
%wendzel@sw-virtMate:~/svn/papers/CUING19_SeqModPaper$ cat detection_results.db | awk -F\; 'BEGIN{acc_sum=0;f1_sum=0;counter=0;}{if ($1 == "4.5"){acc_sum+=$9;f1_sum+=$10;counter+=1;}}END{print "counter=" counter ", accuracy="100*acc_sum/counter, ", f1-score="100*f1_sum/counter;}'
%counter=3, accuracy=94.814 , f1-score=95.0373
%wendzel@sw-virtMate:~/svn/papers/CUING19_SeqModPaper$ cat detection_results.db | awk -F\; 'BEGIN{acc_sum=0;f1_sum=0;counter=0;}{if ($1 == "4.6"){acc_sum+=$9;f1_sum+=$10;counter+=1;}}END{print "counter=" counter ", accuracy="100*acc_sum/counter, ", f1-score="100*f1_sum/counter;}'
%counter=3, accuracy=94.791 , f1-score=95.0347
%wendzel@sw-virtMate:~/svn/papers/CUING19_SeqModPaper$ cat detection_results.db | awk -F\; 'BEGIN{acc_sum=0;f1_sum=0;counter=0;}{if ($1 == "4.7"){acc_sum+=$9;f1_sum+=$10;counter+=1;}}END{print "counter=" counter ", accuracy="100*acc_sum/counter, ", f1-score="100*f1_sum/counter;}'
%counter=3, accuracy=94.1433 , f1-score=94.4563

% 
\begin{table}
 \begin{tabular}{l|cc}
    \toprule
    \textbf{Method} & \textbf{Accuracy} & \textbf{$F_1$ Score} \\
    \midrule
    Threshold, $\kappa=4.4$  & 94.72\% & 94.91\% \\
    Threshold, $\kappa=4.5$  & \textbf{94.81\%} & \textbf{95.04\%} \\
    Threshold, $\kappa=4.6$  & 94.79\% & 95.04\% \\
    \midrule   
%              TP Rate   FP Rate   Precision   Recall  F-Score   ROC Area  Class
%                 0.986     0.046      0.955     0.986     0.97       0.978    CC
%                 0.954     0.014      0.985     0.954     0.969      0.978    NZIX
%Weighted Avg.    0.97      0.03       0.97      0.97      0.97       0.978
%=== Confusion Matrix ===
%
%   a   b   <-- classified as
% 893  13 |   a = CC
%  42 864 |   b = NZIX
    C4.5 Classifier	      &  96.72\% (+1.91\%) & 96.7\% (+1.66\%)      \\
    \bottomrule
 \end{tabular}
\caption{\label{Tab:AllCCMixedResults}Detection results over all covert channels, depending on the threshold.}
\end{table}

\subsection{Further Remarks}

\paragraph*{Implementation:} Our detection approach can either be used on traffic recordings or in a real-time scenario. However, in a \emph{real-time detection scenario}, our approach requires caching of potentially large flows since it uses a stateful algorithm (cf.~\cite[Ch.~8]{NIHbook} and \cite{TrafficNorm}). This renders a real-time detection for huge volumes of data difficult: an attacker could target the detection system by creating a high number of flows that must be cached by the warden, resulting in resource exhaustion, potentially leading to a DoS. However, if limits (e.g.\ on the number of cached flows/unit of time) are enforced, such attacks could be prevented.

On the other hand, if \emph{traffic recordings} are provided and if they are not required to be processed in real-time, our detection approach can be easily used. This fact renders our approach suitable for forensic analyses, e.g.\ under the umbrella of the law-enforcement.

However, a key criterion that we observed during our experiments is the provision of enough RAM if large PCAP files must be processed. Our implementation (non-optimized shell scripts using basically a command pipeline with \texttt{traceconvert}, \texttt{tshark}, \texttt{grep}, \texttt{cut}, \texttt{sed} and \texttt{awk}\footnote{Given enough RAM, an easy way of utilizing multiple cores would be to replace \texttt{grep} with \texttt{xargs}. Our implementation used only one CPU core.}) was capable of handling PCAP files of few hundred MByte on an older SUN Fire X2100 server with 4 GByte of RAM, equipped with an AMD dual core Opteron 180 CPU (running Ubuntu 18.04 LTS) as long as the traffic recordings could be processed entirely in RAM. However, if large traffic recordings must be processed, then all flows and all sequence numbers of each particular flow must be enumerated, which is a computing-intense problem, consuming several days of computing time on our server.\footnote{A more efficient sorting could most likely be achieved if \emph{Apache Hadoop} (or similar distributed solutions) would be applied instead. However, performance was not a concern in our experimental evaluation.} Some NZIX recordings had a compressed size of 2 GByte and we needed to split these files into smaller pieces to process the files on our system (processing these files on a virtual machine with 8 GByte RAM was possible without splitting the files).

\paragraph*{Application by LEAs:} In general, we can assume that our simple heuristic provides good detection results for some of the studied channels. A detection approach using neural networks or similar concepts of AI research could potentially lead to better performing algorithms but also to more opaque ones \cite{OpaqueAlgoInAI}, which can be considered a drawback, especially when explainable, court-proof evidence must be provided by LEA users. For this reason, we can consider the application of our simple algorithm useful for LEAs when the following aspects are taken into account: channels using 2 PDUs were linked to a FPR that is not useful for criminal investigations (5.19\%). Even the small FPRs of 0.6\% and 0.08\% for 3 and 4 PDU channels could provide false evidence for non-criminals due to the chance for false-positives. However, we can assume that it is rather unlikely that only one message ordering flow would be exchanged over a longer period of several days between CS and CR. For this reason, larger traffic volumes should result in several `positives'. However, the lower the compressibility score of a `positive' flow, the more likely a covert channel is present. Thus, for criminal investigations, we recommend not only to determine the number of `positives' but also their compressibility values. Moreover, the \emph{more} positives, the higher the chance that CS and CR did in fact exchange covert information using message ordering.

\section{Conclusion}\label{Sect:Concl}

Covert channels that manipulate the order of PDUs in a network flow appear as legitimate traffic as they do not modify any message's content that would not be modified by legitimate transmissions, too. This quality renders such channels attractive for malware and data exfiltration.

In this paper, we have provided the first countermeasure variation and threshold-based detection approach for such message ordering channels. Our detection method can be applied to every network protocol that can be exploited for a message ordering channel.

Our results have shown that the detectability depends on the number of utilized PDUs of the covert channel and different thresholds should be applied for these channels. 

\emph{Covert channels utilizing 3 or 4 PDUs} could be detected with more than 99.5\% accuracy and $F_1$ score. The low false-positive rate of $<$1\% and $<$0.1\%, respectively, for these channels renders our approach even useful for scenarios with high volumes of traffic. This is important for a real-world application since these two channels can transfer the most information per packet. 
However, \emph{channels utilizing 2 PDUs} could only be detected with an accuracy and $F_1$ score of 94.51\% while their optimal threshold was linked to a false-positive rate of approx.\ 5.1\%, rendering our approach not suitable in practice (except for targeted scenarios with a low traffic volume or when a lower true-positive rate would be acceptable).

In future work, we plan to evaluate whether the so-called $\epsilon$-similarity can lead to higher detection results for the message ordering channels.

\bibliographystyle{acm}
\bibliography{acmart}

\end{document}